\documentclass[twocolumn,showpacs,preprintnumbers,amsmath,amssymb,fleqn]{revtex4}

\usepackage{graphicx}
\usepackage{dcolumn}
\usepackage{bm}

\usepackage{color}
\usepackage{epsfig}

\newcommand{\gtapprox}{\raisebox{-0.5ex}{$\,\stackrel{>}{\scriptstyle\sim}\,$}}
\newcommand{\ltapprox}{\raisebox{-0.5ex}{$\,\stackrel{<}{\scriptstyle\sim}\,$}}

\begin{document}


\preprint{FAU-TP3-08/04 $\quad$ HU-EP-08/18}

\title{Adjoint string breaking in the pseudoparticle approach}

\author{Christian Szasz}

\affiliation{University~of~Erlangen-N{\"u}rnberg, Institute~for~Theoretical~Physics~III, Staudtstra{\ss}e~7, 91058~Erlangen, Germany}

\author{Marc Wagner}

\affiliation{Humboldt~University~Berlin, Department~of~Physics, Newtonstra{\ss}e~15, 12489~Berlin, Germany}

\date{June~11, 2008}

\begin{abstract}

We apply the pseudoparticle approach to SU(2) Yang-Mills theory and perform a detailed study of the potential between two static charges for various representations. Whereas for charges in the fundamental representation we find a linearly rising confining potential, we clearly observe string breaking, when considering charges in the adjoint representation. We also demonstrate Casimir scaling and compute gluelump masses for different spin and parity. Numerical results are in qualitative agreement with lattice results.

\end{abstract}



\maketitle


\section{Introduction}

A common approach to gain some understanding of the vacuum structure of Yang-Mills theory and QCD is to consider effective theories restricting the path integral to a small subset of gauge field configurations, which are supposed to be of physical importance. Well known examples are ensembles of singular gauge instantons (cf.\ \cite{Schafer:1996wv} and references therein), which are able to explain many phenomena on a qualitative level, in particular chiral symmetry breaking. Confinement, however, is absent in these ensembles, although there has been some speculation that very large instantons have the ability to cure this problem (\cite{Diakonov:1996} quoted in \cite{Schafer:1996wv}). More recently related models for Yang-Mills theory have been proposed, which exhibit clear signs of confinement. There are ensembles of regular gauge instantons and merons \cite{Lenz:2003jp,Negele:2004hs,Lenz:2007st}, there is the pseudoparticle approach \cite{Wagner:2005vs,Wagner:2006qn,Wagner:2006du}, there are models of calorons with non-trivial holonomy \cite{Gerhold:2006sk,Gerhold:2006kw} and there is an ensemble of dyons \cite{Diakonov:2007nv}. The successes of these models regarding confinement have either been attributed to the long range nature of their building blocks or to maximally non-trivial holonomy.

However, a satisfactory model for Yang-Mills theory should not only exhibit a linearly rising fundamental potential, but also also Casimir scaling for higher representations as well as $N$-ality dependence. In particular the adjoint potential should exhibit string breaking, when the corresponding charges are separated beyond a certain distance. For a review regarding the confinement problem in Yang-Mills theory we refer to \cite{Greensite:2003bk}.

The goal of this paper is to demonstrate that the pseudoparticle approach applied to SU(2) Yang-Mills theory correctly reproduces the potential between two static charges for various representations. Particular emphasis is put on string breaking in the adjoint representation.


\subsection*{Outline}

When computing the fundamental representation potential, Wilson loops are appropriate observables. In agreement with common expectation the resulting potential is linearly rising both in lattice calculations (cf.\ e.g.\ \cite{Bali:1994de}) and in the pseudoparticle approach (cf.\ e.g.\ \cite{Wagner:2006qn}). In contrast to that, the adjoint potential is expected to saturate at large separations, due to screening by gluons. However, various lattice studies \cite{Michael:1985ne,Griffiths:1985ip,Michael:1991nc,Poulis:1995nn,Stephenson:1999kh,Philipsen:1999wf,deForcrand:1999kr,Kallio:2000jc} have shown that the computation of this potential fails, when using Wilson loops only. One also obtains a linearly rising potential, because of the poor overlap between ``string trial states'' forming the Wilson loops and the ground state, which is essentially a ``two gluelump state'' for large separations (an exception to that is a rather recent study considering 3d Yang-Mills theory, where Wilson loops of extremely large temporal separation have been computed \cite{Kratochvila:2002vm,Kratochvila:2003zj}). A possible way to overcome this problem is to consider whole sets of trial states including both string states (large ground state overlap for small separations) and two gluelump states (large ground state overlap for large separations). The adjoint potential can then be computed from the corresponding correlation matrices via standard variational techniques.

The paper is organized as follows. In section~\ref{SEC_001}, we briefly summarize the basic principle of the pseudoparticle approach, which is discussed in more detail in \cite{Wagner:2006qn}. Then we compute ``pure Wilson loop static potentials'' for charges in various representations (section~\ref{SEC_002}). We determine the fundamental string tension to set the scale, and we demonstrate Casimir scaling for higher representations. In section~\ref{SEC_003} we discuss gluelump creation operators and their quantum numbers. We also present numerical results for a couple of gluelump masses and give estimates regarding the string breaking distance. In section~\ref{SEC_004} we compute the adjoint potential using both string-like and two-gluelump-like trial states. As expected, it is linearly rising at intermediate separations, but saturates at large separations. Moreover, we perform a mixing analysis showing clear evidence for string breaking. Both the string breaking distance and the shape of the potential and its first two excitations are in qualitative agreement with lattice results. Finally we give a summary and a brief outlook (section~\ref{SEC_005}).


\section{\label{SEC_001}The pseudoparticle approach in SU(2) Yang-Mills theory}


\subsection{Introduction to the pseudoparticle approach}

In the following we briefly review the pseudoparticle approach and its application to SU(2) Yang-Mills theory. For a more detailed presentation we refer to \cite{Wagner:2006qn}.

The basic idea of the pseudoparticle approach is to restrict the Yang-Mills path integral to those gauge field configurations, which can be obtained by a linear superposition of a small number of localized building blocks (pseudoparticles). A suitable choice, which is able to reproduce many essential features of SU(2) Yang-Mills theory, particularly a linearly rising fundamental potential, is given by
\begin{eqnarray}
\label{EQN_001} & & \hspace{-0.44cm} a_{\mu,\textrm{instanton}}^b(x) \ \ = \ \ \eta_{\mu \nu}^b \frac{x_\nu}{x^2 + \lambda^2} \\
\label{EQN_002} & & \hspace{-0.44cm} a_{\mu,\textrm{antiinstanton}}^b(x) \ \ = \ \ \bar{\eta}_{\mu \nu}^b \frac{x_\nu}{x^2 + \lambda^2} \\
\label{EQN_003} & & \hspace{-0.44cm} a_{\mu,\textrm{akyron}}^b(x) \ \ = \ \ \delta^{b 1} \frac{x_\mu}{x^2 + \lambda^2} ,
\end{eqnarray}
where $\lambda$ is the pseudoparticle size, \\ $\eta_{\mu \nu}^b = \epsilon_{b \mu \nu} + \delta_{b \mu} \delta_{0 \nu} - \delta_{b \nu} \delta_{0 \mu}$ and \\ $\bar{\eta}_{\mu \nu}^b = \epsilon_{b \mu \nu} - \delta_{b \mu} \delta_{0 \nu} + \delta_{b \nu} \delta_{0 \mu}$. The gauge field configurations entering the path integral are of the form
\begin{eqnarray}
\nonumber & & \hspace{-0.44cm} A_\mu^a(x) \ \ = \ \ \sum_i \mathcal{A}(i) \mathcal{C}^{a b}(i) a_{\mu,\textrm{instanton}}^b(x-z(i)) + \\
\nonumber & & \hspace{0.62cm} \sum_j \mathcal{A}(j) \mathcal{C}^{a b}(j) a_{\mu,\textrm{antiinstanton}}^b(x-z(j)) + \\
\label{EQN_004} & & \hspace{0.62cm} \sum_k \mathcal{A}(k) \mathcal{C}^{a b}(k) a_{\mu,\textrm{akyron}}^b(x-z(k)) ,
\end{eqnarray}
where $z(i) \in \mathbb{R}^4$ denotes the randomly chosen, but fixed position of the $i$-th pseudoparticle, $\mathcal{A}(i) \in \mathbb{R}$ its amplitude and $\mathcal{C}^{ab}(i) \in \textrm{SO(3)}$ its color orientation. Ensemble averages of observables $\mathcal{O}$ are defined by an integration over pseudoparticle degrees of freedom namely amplitudes and color orientations:
\begin{eqnarray}
\Big\langle \mathcal{O} \Big\rangle \ \ = \ \ \frac{1}{Z} \int \left(\prod_i d\mathcal{A}(i) \, d\mathcal{C}(i)\right) \, \mathcal{O}[A] e^{-S[A]} .
\end{eqnarray}
Each gauge field configuration is weighted by $e^{-S}$, where $S$ is the standard Yang-Mills action
\begin{eqnarray}
 & & \hspace{-0.44cm} S \ \ = \ \ \frac{1}{4 g^2} \int d^4x \, F_{\mu \nu}^a F_{\mu \nu}^a \\
 & & \hspace{-0.44cm} F_{\mu \nu}^a \ \ = \ \ \partial_\mu A_\nu^a - \partial_\nu A_\mu^a + \epsilon^{a b c} A_\mu^b A_\nu^c
\end{eqnarray}
with $g$ being the coupling constant. Such finite dimensional integrals can be computed by applying standard Monte Carlo techniques.

In \cite{Wagner:2005vs,Wagner:2006qn,Wagner:2006du} it has been shown that around $400$ pseudoparticles (\ref{EQN_001}) to (\ref{EQN_003}) are sufficient to reproduce many essential features of SU(2) Yang-Mills theory. In particular the potential between static charges in the fundamental representation is linear for large separations. Moreover, like in lattice gauge theory the scale can be set by choosing the coupling constant appropriately. Although the spacetime volume has been varied by a factor of $\approx 16$ with the total number of pseudoparticles kept constant, the dimensionless ratios $\chi^{1/4} / \sigma^{1/2}$ and $T_\textrm{critical} / \sigma^{1/2}$ are essentially independent of the coupling constant and their numerical values are in qualitative agreement with lattice results ($\sigma$: string tension; $\chi$: topological susceptibility; $T_\textrm{critical}$: deconfinement temperature).

Although we use instanton-like building blocks (\ref{EQN_001}) and (\ref{EQN_002}), we would like to stress that the pseudoparticle approach is not a semiclassical model. The intention is rather to approximate physically relevant gauge field configurations with a small number of degrees of freedom. In general these configurations are not close to solutions of the classical Yang-Mills equations of motion.

As has been discussed extensively in \cite{Wagner:2006qn}, the successful qualitative modeling of Yang-Mills physics seems mainly be related to the long range nature of the building blocks (\ref{EQN_001}) to (\ref{EQN_003}) and to the transversality of (\ref{EQN_001}) and (\ref{EQN_002}).


\subsection{Numerical setup}

Unlike in previous applications of the pseudoparticle approach \cite{Wagner:2005vs,Wagner:2006qn,Wagner:2006du} we consider a 4d hypercubic spacetime region (volume: $L^4 = 5.0^4$) with periodic boundary conditions. This allows to fully exploit translational invariance and to adopt certain smearing techniques from lattice gauge theory (cf.\ appendix~\ref{SEC_007} and \ref{SEC_008}). Periodic versions of the pseudoparticles (\ref{EQN_001}) to (\ref{EQN_003}) are obtained by applying a blending technique introduced in \cite{Wagner:2006qn} applied to all four spacetime directions (width of the blending region: $0.25 \times L = 1.25$).

We use $625$ pseudoparticles (\ref{EQN_001}) to (\ref{EQN_003}) with $\lambda = 0.5$. This amounts to a pseudoparticle size, which is of the same order of magnitude as their average nearest neighbor distance. The ratio of pseudoparticles is chosen according to $N_\textrm{instanton}:N_\textrm{antiinstanton}:N_\textrm{akyron} = 3 : 3 : 2$.

We consider values of the coupling constant $g$ between $9.5$ and $18.5$. This corresponds to spacetime extensions in the range of $1.55 \, \textrm{fm} \ltapprox L \ltapprox 2.31 \, \textrm{fm}$ in physical units, where the scale has been set by identifying the fundamental string tension with $4.2 / \textrm{fm}^2$. The majority of computations has been performed at $g = 12.5$, which amounts to $L = 1.85 \, \textrm{fm}$.

All ensemble averages have been computed from $50$ thermalized gauge field configurations (\ref{EQN_004}) with independently chosen pseudoparticle positions. For efficiency of the computation we convert these continuum gauge field configurations to lattice link configurations before ensemble averages are computed (cf.\ appendix~\ref{SEC_006}).


\section{\label{SEC_002}Pure Wilson loop static potentials for different representations}

\begin{figure}[b]
\begin{center}
\includegraphics{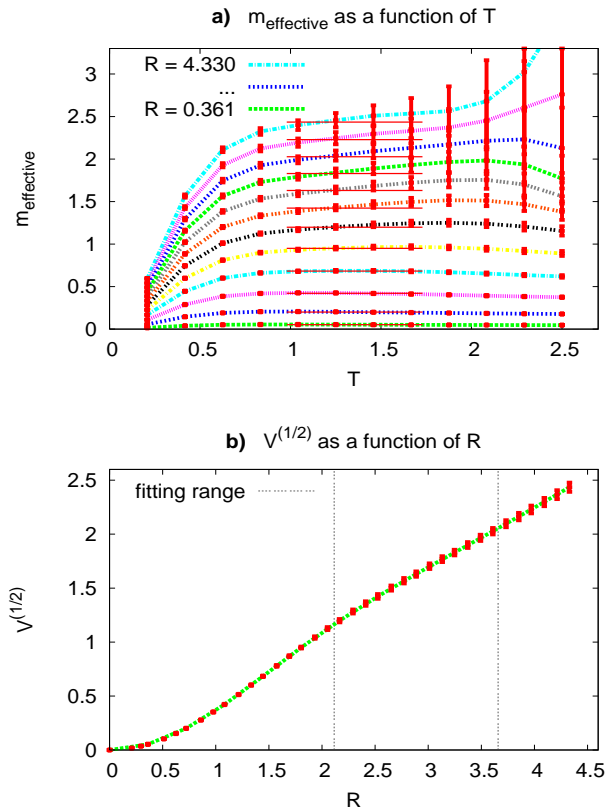}
\caption{\label{FIG_001}
\textbf{a)}~Effective masses for the static potential in the fundamental representation for different separations $R$ as functions of $T$.
\textbf{b)}~The corresponding static potential $V^{(1/2)}$ as a function of the separation $R$.}
\end{center}
\end{figure}

In this section we compute the potential between two static charges for different representations from Wilson loops only. This amounts to considering temporal correlations between string trial states
\begin{eqnarray}
\label{EQN_005} (\phi^{(J)}(\mathbf{x}))^\dagger U^{(J)}(\mathbf{x};\mathbf{y}) \phi^{(J)}(\mathbf{y}) | \Omega \rangle ,
\end{eqnarray}
where $(\phi^{(J)})^\dagger$ and $\phi^{(J)}$ represent static charges in representation $J = 1/2, 1, 3/2, \ldots$ at $\mathbf{x}$ and $\mathbf{y}$ and $U^{(J)}$ is a straight parallel transporter connecting the charges in a gauge invariant way. We orient $U^{(J)}$ along one of the four space diagonals allowing us to consider rather large spatial separations $R = |\mathbf{x}-\mathbf{y}|$ without being affected by periodicity. To maximize the ground state overlap, $U^{(J)}$ is approximated by a product of APE smeared spatial links (cf.\ appendix~\ref{SEC_007}). To compute effective masses, we use the variational method explained in appendix~\ref{SEC_009}. We consider correlation matrices built from three string trial states (\ref{EQN_005}), which differ in their APE smearing parameters ($N_\textrm{APE} \in \{ 5 \, , \, 15 \, , \, 35 \}$, $\alpha_\textrm{APE} = 0.5$).

To give the reader an idea of the plateaux quality obtained from our pseudoparticle computations, we show ground state effective masses for the fundamental representation potential in FIG.~\ref{FIG_001}a. As potential values we take weighted averages of effective masses in ranges where plateaus are indicated (the solid horizontal lines in \\ FIG.~\ref{FIG_001}a). The corresponding potential is plotted in \\ FIG.~\ref{FIG_001}b. As expected it is linear for large separations. For separations smaller than the pseudoparticle size and average nearest neighbor distance cutoff effects are dominant yielding a parabolic rather than a Coulomb-like behavior. To set the scale, we perform a $\chi^2$ minimizing fit with $V(R) = V_0 + \sigma R$ to the data points indicated in FIG.~\ref{FIG_001}b and identify the resulting string tension \\ $\sigma = 0.57(3)$ with the ``physical value'' $4.2 / \textrm{fm}^2$. We obtain a spacetime extension $L = 1.85 \, \textrm{fm}$.

\begin{figure}[t]
\begin{center}
\includegraphics{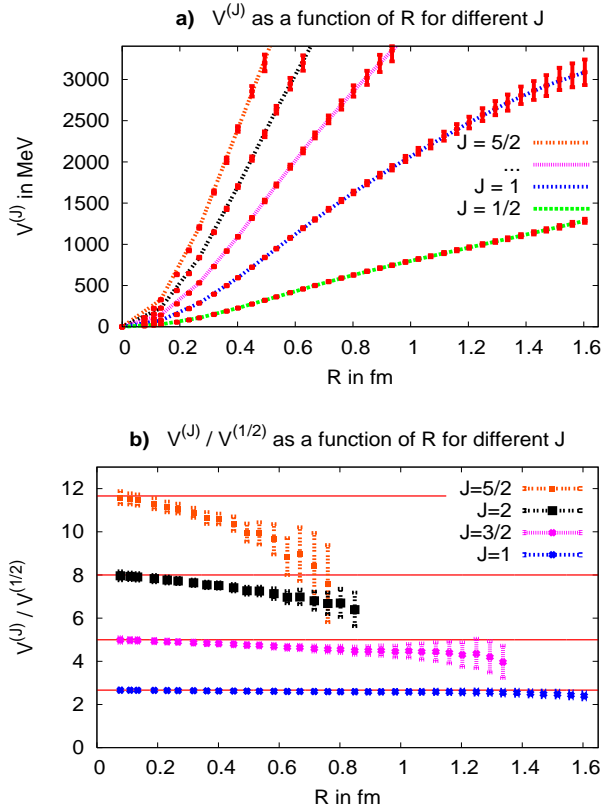}
\caption{\label{FIG_002}
\textbf{a)}~Pure Wilson loop static potentials $V^{(J)}$ for different representations $J$ as functions of the separation $R$.
\textbf{b)}~Ratios of different representation static potentials $V^{(J)} / V^{(1/2)}$ compared to the Casimir scaling expectation as functions of the separation $R$.
}
\end{center}
\end{figure}

It is well known that Wilson loops in higher representations $W_{(R,T)}^{(J)}$, $J \geq 1, 3/2 , 2 , \ldots$, can be expressed in terms of fundamental representation Wilson loops \\ $W_{(R,T)} = W_{(R,T)}^{(1/2)}$ according to
\begin{eqnarray}
 & & \hspace{-0.44cm} W_{(R,T)}^{(1)} \ \ = \ \ \frac{4}{3} \Big(W_{(R,T)}\Big)^2 - \frac{1}{3} \\
 & & \hspace{-0.44cm} W_{(R,T)}^{(3/2)} \ \ = \ \ 2 \Big(W_{(R,T)}\Big)^3 - W_{(R,T)} \\
 & & \hspace{-0.44cm} W_{(R,T)}^{(2)} \ \ = \ \ \frac{16}{5} \Big(W_{(R,T)}\Big)^4 - \frac{12}{5} \Big(W_{(R,T)}\Big)^2 + \frac{1}{5} \\
\nonumber & & \hspace{-0.44cm} W_{(R,T)}^{(5/2)} \ \ = \ \ \frac{16}{3} \Big(W_{(R,T)}\Big)^5 - \frac{16}{3} \Big(W_{(R,T)}\Big)^3 + W_{(R,T)} . \\
 & &
\end{eqnarray}
Results for the corresponding potentials are shown in FIG.~\ref{FIG_002}a. In FIG.~\ref{FIG_002}b we plot ratios $V^{(J)} / V^{(1/2)}$. For intermediate distances these ratios are expected to be close to the ratios of the corresponding Casimir operators, which are given by $8/3$ ($J=1$), $5$ ($J=3/2$), $8$ ($J=2$) and $35/3$ ($J=5/2$) (the solid lines in FIG.~\ref{FIG_002}b). While the adjoint potential is in excellent agreement with the Casimir scaling hypothesis, higher representations exhibit certain deviations at larger separations. These findings are in agreement with both \cite{Lenz:2007st} and \cite{Piccioni:2005un}, where similar analyses in related meron and regular gauge instanton ensembles as well as in 4d SU(2) lattice gauge theory have been performed.

Note that there is no sign of string breaking in the adjoint potential ($J=1$ curve in FIG.~\ref{FIG_002}a), which is expected to happen at distances $R \approx 1.0 \, \textrm{fm} \, \ldots \, 1.25 \, \textrm{fm}$ \cite{Jorysz:1987qj,deForcrand:1999kr}. As we will demonstrate in section~\ref{SEC_004} this is due to the poor ground state overlap of string trial states (\ref{EQN_005}) for separations larger than the string breaking distance.


\section{\label{SEC_003}Gluelump masses for different spin and parity}

To estimate the string breaking distance, we compute masses of states containing a single static adjoint charge surrounded by gluons, so called gluelumps.


\subsection{Gluelump trial states}

The symmetry group of states constrained by a single static charge $\phi^{(1)}$ located at position $\mathbf{x}$ is $SO(3) \times P$ (rotation and parity both with respect to $\mathbf{x}$). Suitable ``gluelump creation operators'' with well defined quantum numbers are given by
\begin{eqnarray}
\label{EQN_006} & & \hspace{-0.44cm} G_j^{(J=1,P=+)}(\mathbf{x}) \ \ = \ \ \textrm{Tr}\Big(\Phi(\mathbf{x}) B_j(\mathbf{x})\Big) \\
\label{EQN_007} & & \hspace{-0.44cm} G_j^{(J=1,P=-)}(\mathbf{x}) \ \ = \ \ \textrm{Tr}\Big(\Phi(\mathbf{x}) \epsilon_{j k l} D_k B_l(\mathbf{x})\Big) \\
\label{EQN_008} & & \hspace{-0.44cm} G_j^{(J=2,P=-)}(\mathbf{x}) \ \ = \ \ \textrm{Tr}\Big(\Phi(\mathbf{x}) |\epsilon_{j k l}| D_k B_l(\mathbf{x})\Big) ,
\end{eqnarray}
where $j \in \{x \, , \, y \, , \, z\}$, $\Phi = (\phi^{(1)})^a \sigma^a / 2$, $B_j$ denotes the color magnetic field and $D_j = \partial_j - i [A_j,\ldots]$ the covariant derivative. In the literature $(J=1,P=+)$ and \\ $(J=1,P=-)$ gluelumps are also referred to as magnetic and electric gluelumps respectively.

Since we perform a latticization of our pseudoparticle gauge field configurations, it is convenient to replace $B_j$ by the magnetic clover leaf in a plane perpendicular to the $j$-direction and $D_k B_l$ by the corresponding electric clover leaf in a plane perpendicular to the $l$-direction \cite{Michael:1985ne}. After this replacement our gluelump creation operators are not rotationally symmetric anymore, but belong to one of the irreducible representations of the cubic rotation group $\textrm{O}_\textrm{h}$: $T_1$ for (\ref{EQN_006}) and (\ref{EQN_007}), i.e.\ the corresponding states are superpositions of spin $J = 1,3,4,\ldots$, and $T_2$ for (\ref{EQN_008}), i.e.\ spin values $J = 2,3,4,\ldots$ are possible.


\subsection{Numerical results}

We determine gluelump masses from temporal correlations
\begin{eqnarray}
\nonumber & & \hspace{-0.44cm} \langle \Omega | \Big(G_j^{(J,P)}(\mathbf{x},T)\Big)^\dagger G_j^{(J,P)}(\mathbf{x},0) | \Omega \rangle \quad \textrm{(no sum over }j\textrm{)} , \\
 & &
\end{eqnarray}
where the spatial links appearing in the clover leafs are APE smeared, to maximize the ground state overlap \\ ($N_\textrm{APE} = 5$, $\alpha_\textrm{APE} = 0.5$; cf.\ appendix~\ref{SEC_007} and \ref{SEC_009}).

\begin{figure}[t]
\begin{center}
\includegraphics{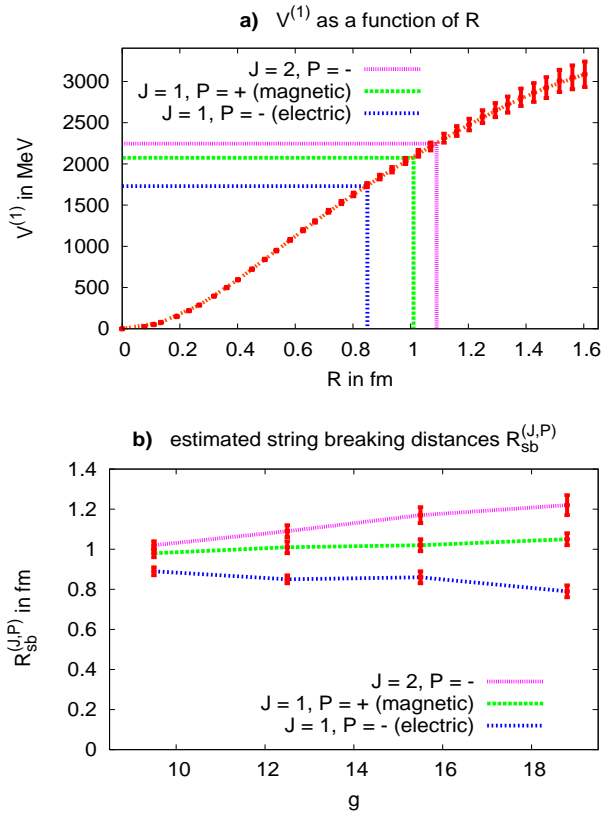}
\caption{\label{FIG_003}
\textbf{a)}~Estimating string breaking distances $R_\textrm{sb}^{(J,P)}$ by intersecting the static adjoint potential $V^{(1)}$ with $2 \, m_\textrm{gluelump}^{(J,P)}$.
\textbf{b)}~Estimated string breaking distances $R_\textrm{sb}^{(J,P)}$ as functions of the coupling constant $g$ for different gluelump quantum numbers $(J,P)$.
}
\end{center}
\end{figure}

Note that gluelump masses by themselves are not physically meaningful. They are cutoff dependent quantities, which diverge in the continuum limit, due to the self energy of the static charge (cf.\ e.g.\ \cite{Laine:1997nq,Bali:2003jq}). However, mass differences of gluelumps or more generally of states containing the same number of static charges are physical observables. This offers the possibility to compute mass differences between gluelumps and to estimate the string breaking distance $R_\textrm{sb}^{(J,P)}$ by intersecting the pure Wilson loop adjoint potential $V^{(1)}$ with two times the corresponding gluelump mass (cf.\ FIG.~\ref{FIG_003}a). Results are collected in TABLE~\ref{TAB_001}.

\begin{table}[b]
\begin{tabular}{|c|c|c|c|}
\hline
 & & & \vspace{-0.25cm} \\
$(J,P)$ & $m_\textrm{gluelump}^{(J,P)}$ & $m_\textrm{gluelump}^{(J,P)} - m_\textrm{gluelump}^{(1,-)}$ & $R_\textrm{sb}^{(J,P)}$ \\
 & & & \vspace{-0.25cm} \\
\hline
 & & & \vspace{-0.25cm} \\
$(1,+)$ & $1037(21) \, \textrm{MeV}$ & $172(38) \, \textrm{MeV}$ & $1.01(3) \, \textrm{fm}$ \\
 & & & \vspace{-0.25cm} \\
$(1,-)$ & $865(17) \, \textrm{MeV}$ & $-$ & $0.85(2) \, \textrm{fm}$ \\
 & & & \vspace{-0.25cm} \\
$(2,-)$ & $1123(23) \, \textrm{MeV}$ & $257(40) \, \textrm{MeV}$ & $1.09(3) \, \textrm{fm}$\vspace{-0.25cm} \\
 & & & \\
\hline
\end{tabular}
\caption{\label{TAB_001}Gluelump masses, mass differences and estimated string breaking distances for different quantum numbers $(J,P)$.}
\end{table}

Comparing these results to available lattice results we find rather good agreement for the estimated ``magnetic string breaking distance'': $R_\textrm{sb,lattice}^{(1,+)} = 1.0 \, \textrm{fm} \, ... \, 1.25 \, \textrm{fm}$ \cite{deForcrand:1999kr,Jorysz:1987qj}. Moreover, the mass of the magnetic and of the electric gluelump are of the same order of magnitude. However, their mass difference, which is hard to measure precisely, due to large absolute mass values, has the opposite sign: $(m_\textrm{gluelump}^{(1,+)} - m_\textrm{gluelump}^{(1,-)})_\textrm{lattice} = -203(76) \, \textrm{MeV}$ \cite{Jorysz:1987qj}. An explanation might be that ultraviolet fluctuations, which are probably more important for localized objects like gluelumps than for the static potential at large separations, are not adequately described by our pseudoparticle regularization (the resolution of our gauge field configurations is roughly given by the pseudoparticle size and the average nearest neighbor distance, which is $\approx 0.4 \, \textrm{fm}$ for $g = 12.5$). Similar problems have been encountered in a study of the gluelump spectrum in related meron and regular gauge instanton ensembles \cite{Lenz:2007st}.

We have also investigated the stability of the estimated string breaking distances under a  variation of the coupling constant $g$ (cf.\ FIG.~\ref{FIG_003}b). While the spacetime volume in physical units has been increased by a factor of \\ $\approx 5$ ($g=9.5 \ldots 18.5$ corresponds to \\ $L = 1.55 \, \textrm{fm} \, \ldots \, 2.31 \, \textrm{fm}$) with the total number of pseudoparticles kept constant, the estimated string breaking distances $R_\textrm{sb}^{(1,+)}$, $R_\textrm{sb}^{(1,-)}$ and $R_\textrm{sb}^{(2,-)}$ vary by only $\approx 6 \%$, $\approx 9 \%$ and $\approx 19 \%$ respectively. This complements the scaling analysis performed in \cite{Wagner:2006du}, where the dimensionless ratios $\chi^{1/4} / \sigma^{1/2}$ and $T_\textrm{critical} / \sigma^{1/2}$ have been found to be essentially independent of the coupling constant.


\section{\label{SEC_004}Adjoint string breaking}

In this section we supplement our basis of string trial states (\ref{EQN_005}) used in the variational method (cf.\ appendix~\ref{SEC_009}) by a second type of trial state, which resembles a two-gluelump state.


\subsection{Two-gluelump trial states}

When computing the adjoint potential, it is essential to have a basis of trial states with significant ground state overlap for arbitrary separations of the static charges. For intermediate separations the ground state is expected to be a gluonic string connecting the charges, i.e.\ a state with large overlap to the string trial states (\ref{EQN_005}). For large separations the charges are supposed to be screened by gluons forming two essentially non-interacting gluelumps. As we will demonstrate at the end of this section, the overlap to string trial states is rather poor.

In the following we consider the static charges $(\phi^{(1)})^\dagger$ and $\phi^{(1)}$ located at positions $\mathbf{x} = (0,0,-z/2)$ and \\ $\mathbf{y} = (0,0,+z/2)$. The corresponding symmetry group is $\mathrm{SO(2)} \times P_z (\times P_x)$. The $\mathrm{SO(2)}$ rotation is around the $z$-axis with angular momentum $J_z$ as corresponding quantum number. $P_z$ denotes reflection along the $z$-axis with respect to its center, and in the case of $J_z = 0$ there is another symmetry $P_x$, reflection along the $x$-axis \cite{Griffiths:1983ah}.

The quantum numbers of the string trial states (\ref{EQN_005}) are $J_z = 0$, $P_z = +$ and $P_x = +$. Since we are not only interested, whether the adjoint potential saturates at two times the gluelump mass, but also whether the string actually breaks, when static charges are separated adiabatically, we need two-gluelump trial states, which have the same quantum numbers. Using products of two single-gluelump creation operators (\ref{EQN_006}) to (\ref{EQN_008}) an obvious choice is
\begin{eqnarray}
\nonumber & & \hspace{-0.44cm} \Big(G_x^{(J,P)}(\mathbf{x}) G_x^{(J,P)}(\mathbf{y}) + G_y^{(J,P)}(\mathbf{x}) G_y^{(J,P)}(\mathbf{y}) + \\
\label{EQN_009} & & \hspace{0.62cm} G_z^{(J,P)}(\mathbf{x}) G_z^{(J,P)}(\mathbf{y})\Big) | \Omega \rangle .
\end{eqnarray}
There is another possibility, where $G_z^{(J,P)} G_z^{(J,P)}$ is weighted by a factor of $-2$. However, since string breaking is closely related to non-vanishing off-diagonal elements in correlation matrices, we expect such a state to be less suited for our purposes, due to the relative minus sign. Therefore, we supplement our basis of string trial states by (\ref{EQN_009}) for the following computations.


\subsection{Numerical results}

Although the electric gluelump turned out to be lighter than the magnetic gluelump, we compute string breaking both with electric and with magnetic trial states. This allows a direct comparison to lattice results, since there seem to be only investigations of 4d adjoint string breaking with magnetic trial states in the literature \cite{deForcrand:1999kr,Kallio:2000jc}.

\begin{figure}[b]
\begin{center}
\includegraphics{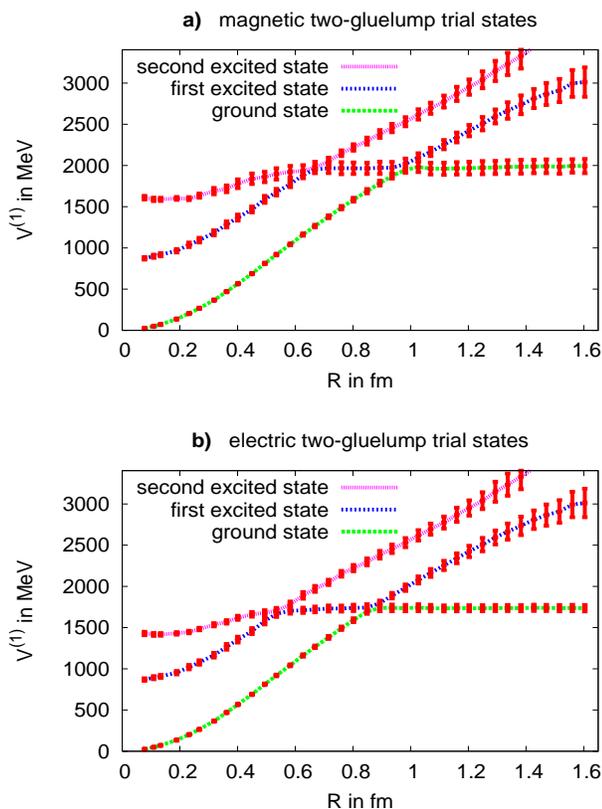}
\caption{\label{FIG_004}The static adjoint potential $V^{(1)}$ and its first two excitations as functions of the separation $R$.
\textbf{a)}~Magnetic two-gluelump trial state.
\textbf{b)}~Electric two-gluelump trial state.
}
\end{center}
\end{figure}

We extract the adjoint potential from $3 \times 3$ correlation matrices containing two string trial states \\ ($N_\textrm{APE} \in \{ 15 \, , \, 35 \}$, $\alpha_\textrm{APE} = 0.5$; cf.\ appendix~\ref{SEC_007}) and one two-gluelump-trial state ($N_\textrm{APE} = 5$, $\alpha_\textrm{APE} = 0.5$) by means of the variational technique explained in appendix~\ref{SEC_009}. Results, which are shown in FIG.~\ref{FIG_004}, are qualitatively the same both for magnetic and electric trial states. In contrast to the pure Wilson loop computation (cf.\ FIG.~\ref{FIG_002}a, $J=1$ curve) the potential saturates at two times the gluelump mass and at separations close to the estimated string breaking distance (cf.\ TABLE~\ref{TAB_001}). We also plot the first and second excitation. It is interesting to note that for small separations the first excitation is a string excitation, for intermediate distances it becomes a two gluelump state and for large separations it is a string state again. This level ordering is in agreement with various lattice computations in 3d \cite{Stephenson:1999kh,Philipsen:1999wf} and 4d \cite{deForcrand:1999kr} SU(2) Yang-Mills theory. Moreover, within $\approx 20\%$ there is agreement with \cite{deForcrand:1999kr} regarding the string breaking distance and the separation of the energy levels.

\begin{figure}[b]
\begin{center}
\includegraphics{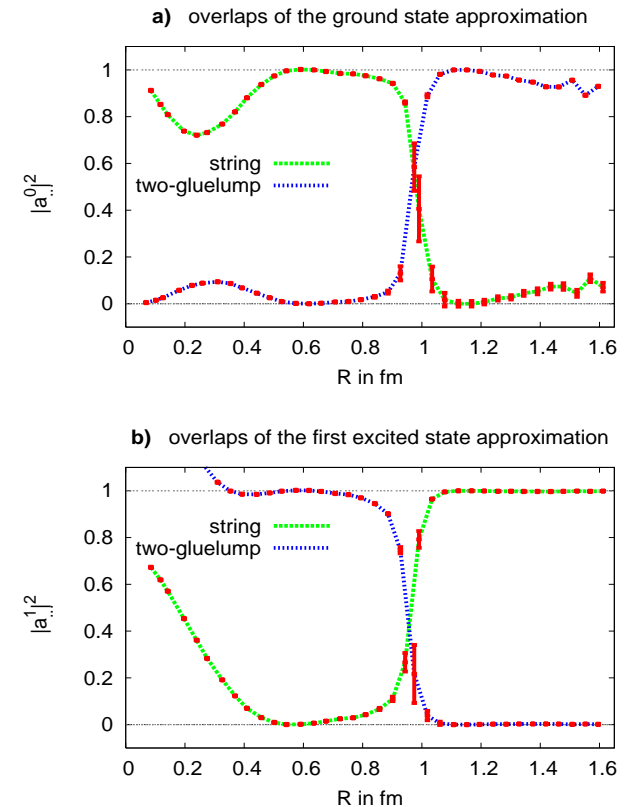}
\caption{\label{FIG_005}
\textbf{a)}~Overlaps of the ground state approximation to the trial states as functions of the separation $R$.
\textbf{b)}~Overlaps of the first excited state approximation to the trial states as functions of the separation $R$.
}
\end{center}
\end{figure}

We have also investigated, whether the string actually breaks, when static charges are separated adiabatically, or whether there is just a plain level crossing of string and two-gluelump ground states. Since this question is rather hard to resolve from potential plots like FIG.~\ref{FIG_004}, we perform a mixing analysis in close analogy to \cite{deForcrand:1999kr}. To keep things as simple as possible, we consider two normalized trial states, a string trial state $| \textrm{string} \rangle$ ($N_\textrm{APE} = 35$, $\alpha_\textrm{APE} = 0.5$) and a two gluelump-trial state $| \textrm{two-gluelump} \rangle$ ($N_\textrm{APE} = 5$, $\alpha_\textrm{APE} = 0.5$). From the variational method we obtain approximations of the ground and the first excited state:
\begin{eqnarray}
\nonumber & & \hspace{-0.44cm} | 0 \rangle \ \ \approx \ \ a_\textrm{string}^0 | \textrm{string} \rangle + a_\textrm{two-gluelump}^0 | \textrm{two-gluelump} \rangle \\
 & & \\
\nonumber & & \hspace{-0.44cm} | 1 \rangle \ \ \approx \ \ a_\textrm{string}^1 | \textrm{string} \rangle + a_\textrm{two-gluelump}^1 | \textrm{two-gluelump} \rangle . \\
& &
\end{eqnarray}
In FIG.~\ref{FIG_005} we show the squared amplitudes $|a_\textrm{string}^j|^2$ and $|a_\textrm{two-gluelump}^j|^2$ as functions of the separations for the case of a magnetic two-gluelump trial state. It is clearly visible that for separations smaller than the estimated string breaking distance $R_\textrm{sb}^{(1,+)} \approx 1.0 \, \textrm{fm}$ the ground state is essentially a string state, while the first excited state is almost exclusively a two-gluelump state. For separations larger than the string breaking distance the situation is reversed, explaining why a computation of string breaking from Wilson loops alone has failed. In a narrow range around the string breaking distance we observe mixing of string and two gluelump trial states. This mixing indicates that there is a smooth transition from a string state for $R \ltapprox R_\textrm{sb}^{(1,+)}$ to a two gluelump state for $R \gtapprox R_\textrm{sb}^{(1,+)}$, when static charges are separated adiabatically. We conclude that string breaking is present in the pseudoparticle approach. Comparing the overlap plots from FIG.~\ref{FIG_005} to the lattice result in \cite{deForcrand:1999kr} we find again rather good agreement.


\section{\label{SEC_005}Summary}

We have performed a detailed study of adjoint string breaking in the pseudoparticle approach. In agreement with lattice gauge theory the static potential saturates at two times the gluelump mass, which corresponds for magnetic trial states to a charge separation of $\approx 1.0 \, \textrm{fm}$. Moreover, from a mixing analysis we have obtained strong indications that the connecting gluonic string actually breaks, when the corresponding charges are separated adiabatically. We have also computed static potentials for various representations from Wilson loops only. There is excellent agreement with the Casimir scaling hypothesis for the adjoint potential and also higher representations exhibit only minor deviations. In view of these successes we conclude that the pseudoparticle approach is a model for SU(2) Yang-Mills theory, which correctly reproduces many essential features connected to confinement.

Gluelump masses, on the other hand, differ to some extent from lattice results. In particular the level ordering of the magnetic and the electric gluelump is reversed. This might indicate the limitations of pseudoparticle regularizations, where degrees of freedom are chosen to model long range correlations rather than ultraviolet fluctuations.


\section*{Outlook}

After this successful computation of screening of adjoint charges within the pseudoparticle approach, a natural next step is an investigation of string breaking in QCD (for a recent lattice study cf.\ \cite{Bali:2005fu}). First steps regarding the treatment of fermionic fields in terms of pseudoparticles have already been taken \cite{Wagner:2007he,Wagner:2007av}. An interesting feature of the pseudoparticle approach regarding numerical efficiency is the fact that it uses a rather small number of degrees of freedom compared to lattice gauge theory. This makes exact computations of fermionic all-to-all propagators not only possible, but also extremely cheap. This might offer the possibility to compute fermionic quantities on a qualitative level without using high performance computer resources.


\begin{acknowledgments}

We would like to thank Martin Ammon, Hartmut Hofmann, Ernst-Michael Ilgenfritz, Frieder Lenz and Michael M\"uller-Preussker for helpful discussions and useful comments. 

\end{acknowledgments}


\appendix

\section{\label{SEC_006}Latticization of pseudoparticle gauge field configurations}

Before computing ensemble averages we convert the continuum gauge field configurations (\ref{EQN_004}) to $24^4$ lattice link configurations. We do this to increase the efficiency of our computations and to adopt certain smearing techniques from lattice gauge theory. The corresponding links are computed by sampling (\ref{EQN_004}) sufficiently often along these links and by multiplying corresponding SU(2) matrices. We would like to stress that, although ensemble averages are computed on such lattices, the underlying gauge field configurations are still continuum gauge field configurations. Therefore, all results presented in this paper are results from a continuum model, where the lattice has only been introduced for the sake of convenience and numerical efficiency.


\subsection{\label{SEC_007}APE smearing of spatial links}

The ground state overlaps of string, gluelump and two-gluelump trial states (eqns.\ (\ref{EQN_005}), (\ref{EQN_006}) to (\ref{EQN_008}) and (\ref{EQN_009})) can be increased by giving the corresponding creation operators certain volume extensions. Such operators can be obtained by replacing all spatial links $U_j = U_j^{(0)}$ by their APE smeared versions $U_j^{(N_\textrm{APE})}$:
\begin{eqnarray}
  \nonumber & & \hspace{-0.44cm} U_j^{(N+1)}(x) \ \ = \ \ P_\textrm{SU(2)}\bigg(U_j^{(N)}(x) + \alpha_\textrm{APE} \sum_{k=\pm 1,\pm 2,\pm 3}^{k \neq \pm j} \\
 & & \hspace{0.62cm} U_k^{(N)}(x) U_j^{(N)}(x+e_k) U_{-k}^{(N)}(x+e_k+e_j)\bigg) ,
\end{eqnarray}
where $P_\textrm{SU(2)}$ denotes an appropriate normalization projecting back to $\textrm{SU(2)}$ \cite{Albanese:1987ds}.


\subsection{\label{SEC_008}HYP smearing of temporal links}

To reduce the self energy of static charges, which in turn significantly improves the signal-to-noise ratio, we use HYP smearing of temporal links \cite{Hasenfratz:2001hp}. HYP smearing is reminiscent to three iterations of APE smearing, where links outside a hypercube around the original link are ignored. There are three parameters, which have been ``optimized'' in \cite{Della Morte:2005yc} and are commonly referred to as HYP2: $\vec{\alpha}_\textrm{HYP2} = (1.0 \, , \, 1.0 \, , \, 0.5)$.

\begin{figure}[h]
\begin{center}
\includegraphics{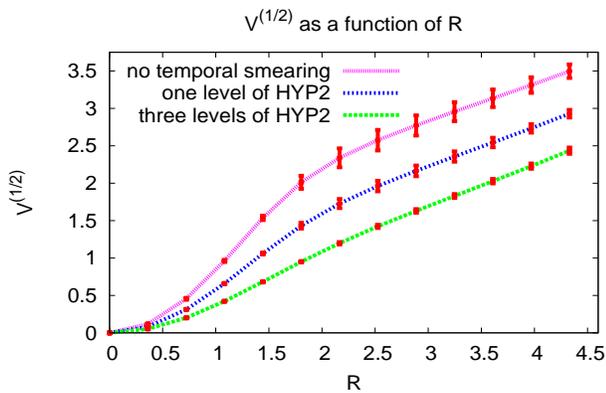}
\caption{\label{FIG_006}The static potential in the fundamental representation $V^{(1/2)}$ as a function of the separation $R$ (unsmeared temporal links, one level of HYP2 and three levels of HYP2).}
\end{center}
\end{figure}

Throughout this paper we use three iterations of HYP2 smearing for temporal links. Since the extension of the resulting links is still below the pseudoparticle cutoff, which is of the order of the pseudoparticle size and the nearest neighbor distance, we do not expect to alter correlation functions at separations, where physically meaningful results can be extracted. We checked that this is indeed the case for various observables. An example, the static potential in the fundamental representation, is shown in FIG.~\ref{FIG_006}. It is obvious that the slope of the potential for large separations is the same for unsmeared and for HYP2 smeared temporal links, while statistical errors are significantly reduced for the latter.

\section{\label{SEC_009}Correlation matrices and effective masses}

To determine gluelump masses and static potentials more reliably, we use a well known variational technique (cf.\ e.g.\ \cite{Griffiths:1983ah}).

The starting point is a correlation matrix
\begin{eqnarray}
C_{J K}(T) \ \ = \ \ \langle \Omega | \Big(\mathcal{O}_J(T)\Big)^\dagger \mathcal{O}_K(0) | \Omega \rangle ,
\end{eqnarray}
where $\mathcal{O}_J$ are suitable creation operators yielding a trial state basis of string, gluelump or two gluelump states with appropriate quantum numbers and possibly different extensions. The basis should be chosen such that a good approximation of the ground state is possible.

Once this correlation matrix has been computed, one has to solve the generalized eigenvalue problem
\begin{eqnarray}
C_{J K}(T_0) v_K^{(N)}(T_0) \ \ = \ \ C_{J K}(T_0-a) v_K^{(N)}(T_0) \lambda^{(N)}
\end{eqnarray}
at a fixed value of $T_0$. Approximations of low lying states $| N \rangle$ are given by
\begin{eqnarray}
 & & \hspace{-0.44cm} | N \rangle \ \ \approx \ \ v_K^{(N)} \mathcal{O}_K | \Omega \rangle 
%
\end{eqnarray}
and corresponding energies can be determined from effective mass plateaus:
\begin{eqnarray}
\nonumber & & \hspace{-0.44cm} m_\textrm{effective}^{(N)}(T) \ \ = \\
 & & = \ \ -\frac{1}{a} \ln\bigg(\frac{(v_J^{(N)}(T_0))^\ast C_{J K}(T) v_K^{(N)}(T_0)}{(v_J^{(N)}(T_0))^\ast C_{J K}(T-a) v_K^{(N)}(T_0)}\bigg) .
\end{eqnarray}



\end{document}